\documentclass[aps,prl,twocolumn,groupedaddress,showpacs,letterpaper]{revtex4}
\usepackage{amsmath}
\usepackage{graphicx}

\begin{document}

\title{Chirality-Dependent Growth Rate of Carbon Nanotubes - A Theoretical Study}

\author{Heiko Dumlich\ref{aff:FUBerlin}}
\email[Corresponding author: ]{heiko.dumlich@fu-berlin.de}

\author{Stephanie Reich\ref{aff:FUBerlin}}

\affiliation{\label{aff:FUBerlin} Fachbereich Physik, Freie Universit\"{a}t Berlin,
14195 Berlin, Germany}

\date{\today}

\begin{abstract}
We consider geometric constraints for the addition of carbon atoms 
to the rim of a growing nanotube. The growth of a tube proceeds 
through the conversion of dangling bonds from armchair to zigzag 
and vice versa. We find that the growth rate depends on the rim 
structure (chirality), the energy barriers for dangling bond 
conversion, and the growth temperature. A calculated chirality 
distribution derived from this minimalistic theory shows surprisingly 
good agreement with experiment. Our ideas imply that the chirality 
distribution of carbon nanotubes can be influenced by external 
parameters.
\end{abstract}


\maketitle

\section{I. Introduction}
The properties of carbon nanotubes~\cite{Iijima1993a} depend strongly 
on their chirality or atomic structure. Most notably, the metallic and 
semiconducting character and the band gap of a tube change with $\left(n,m\right)$ 
chiral index. One of the greatest challenges in nanotube research and 
application is to control the chirality during the growth. This would 
allow the production of tubes with tailored properties without relying 
on a sorting of bulk samples.

The growth of a nanotube can conceptually be divided into two stages: 
the nucleation of a cap and the elongation of the nucleus into a tube
~\cite{H.Amara2008,Ohta2009,Jean-YvesRaty2005,X.Fan2003,Yazyev2008}. 
Reich \emph{et al.}~\cite{StephanieReich2006} showed that the nucleation 
of the cap fixes the chirality of an individual tube as a change in 
chirality is unlikely during the growth phase. Harutyunyan 
\emph{et al.}~\cite{AvetikR.Harutyunyan2009} reported preferential 
growth of metallic tubes and claimed the selection to follow from the 
shape of the catalytic particles, i.e., chirality selection during the 
nucleation phase.

The final volume fraction of a given nanotube type does not only depend 
on the nucleation, but also on growth speed during elongation. Elongation 
was mainly studied in simplistic models with carbon addition
~\cite{Gomez-Gualdron2008,Qiang2010}. Ding \emph{et al.}~\cite{FengDing2009} 
argued that achiral armchair and zigzag tubes grow by introducing kinks 
when starting a new layer. They predicted the armchair kinks to require 
much less energy than zigzag kinks. The growth process, which is driven 
by a monotonous decrease in free energy during elongation, will, therefore, 
favour armchair tubes. Within this line of reasoning chirality selection is 
independent of external parameters such as catalyst type and temperature.

In this paper, we suggest that the chiral-angle distribution of carbon 
nanotubes depends on external parameters. The key is to manipulate the 
energy difference between armchair and zigzag dangling bonds through the 
choice of metal catalyst and growth conditions. We arrive at this conclusion 
by looking at the geometry of a growing tube, the number and types of places 
for carbon addition. The rim of a nanotube consists of three different growth 
sites with a varying energy barrier for the addition of carbon atoms. The 
number and distribution of growth sites is a function of chirality. Combining 
this minimalistic geometric approach with calculated energy differences for 
carbon dangling bonds on metals we predict a distribution of chiral angles 
that is in surprisingly good agreement with experimental findings.

This paper is organized as follows. We first show how growth proceeds with 
carbon addition with respect to our model, Sec. II A. The essential properties 
of rims made up by hexagons are discussed in Sec. II B. The growth factor, which 
allows us to understand why chiral selectivity occurs during the nanotube elongation, 
is introduced in Sec. II C. We then discuss how the chirality distribution can 
be influenced by external parameters in Sec. II D. Finally, in Sec. III we use 
our model - derived in Sec. II - to obtain two exemplary chirality distributions 
and compare our results to experimentally determined data. Section IV summarizes 
this work.

\section{II. Model}
\subsection{A. Growth}
\begin{figure*}
\includegraphics[scale=0.18]{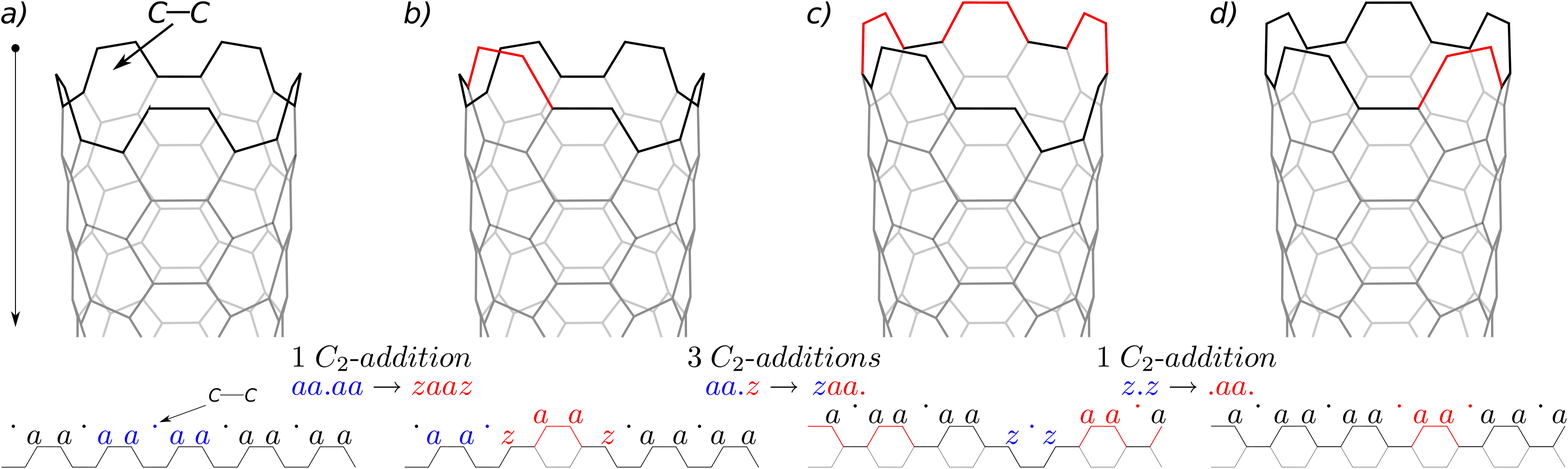}

\caption{\label{fig:fig1}Three dimensional (top) and reduced rim (bottom)
representations of half layer growth of a $\left(5,5\right)$ carbon
nanotube. The reduced rim representations at the bottom are obtained
by unzipping the 3d wire model (top). The {}``.'' denotes a growth
site for $\mbox{C}_{2}$ addition, the $a$ stands for an armchair
and the $z$ for a zigzag dangling bond. The arrow at the left denotes
the growth direction. We assume root growth, but the picture is turned
upside down and a catalyst was omitted for clarity of the $\mbox{C}_{2}$
addition. \emph{a)} A $\mbox{C}_{2}$ adds at the rim. The induction
of a new layer is accompanied by a barrier~\cite{FengDing2009}.\emph{
b)} 3 $\mbox{C}_{2}$ additions follow without experiencing an energy
barrier leading to tube \emph{c)}. The last $\mbox{C}_{2}$ addition
leads to a stable rim or closed layer. \emph{d)} A half layer is grown
compared to \emph{a)}. This process continues until the growth is
terminated.}
 
\end{figure*}

Let us first consider schematically the growth of a carbon nanotube.
In Fig.~\ref{fig:fig1} we present a 3d wire model of a possible growth
route of a $\left(5,5\right)$ nanotube. The growth proceeds through
to addition of $\mbox{C}_{2}$. The first carbon atom adds endothermically
and is followed exothermically by a second carbon atom. The pentagon 
created in the first step is energetically less favorable than hexagons. 
We, therefore, expect the next carbon atom to be added to the pentagon. 
Also, the creation of more and more pentagons would close the tube and 
terminate growth~\cite{YoungHeeLee1997}. Alternatively, a carbon dimer 
is added.

Going through the series of tubes in Fig.~\ref{fig:fig1}\emph{a)}
to \emph{d)} a layer of carbon atoms was grown, which corresponds
to half a unit cell of the $\left(5,5\right)$ tube. The continuation
of the process - until growth is terminated - leads to an armchair
carbon nanotube. In the lower panels of Fig.~\ref{fig:fig1}\emph{a)}
to \emph{d)} we introduce a schematic representation of the growing
rim of a nanotube. The reduced rim representation unfolds the rim
of the tubes in two dimensions. The {}``.'' denote growth sites
at which the addition of $\mbox{C}_{2}$ is energetically favorable
because a hexagon is created. The $a$ stands for an armchair dangling
bond; it consists of one of two neighbouring twofold C-C bonded atoms.
The $z$ stands for zigzag dangling bond; it has two saturated C neighbours
and is itself twofold C-C bonded.

The first $\mbox{C}_{2}$ addition to the rim starts a new layer by
converting two $a$ into $z$ dangling bonds, which is accompanied
by an energy barrier~\cite{FengDing2009}, compare Fig.~\ref{fig:fig1}\emph{a)}
and \emph{b)}. The following three $\mbox{C}_{2}$ additions do not
change the energy of the rim, since they only move $a$ and $z$ dangling
bonds leading to Fig.~\ref{fig:fig1}\emph{c)}. The final $\mbox{C}_{2}$
addition to Fig.~\ref{fig:fig1}\emph{c)} yields a finished armchair
layer presented in \emph{d)}. We argue that the conversion and movement
of the growth sites and the energetic barriers for the conversion
determine the chirality-specific growth speed of carbon nanotubes.

\subsection{B. Rim}
In the following we first discuss the essential properties of rims
made up by hexagons. All carbon nanotube rims consist of armchair
$a$ and zigzag-type dangling bonds $z$. The number and nature of
the dangling bonds in a rim depend mainly on chirality $\left(n,m\right)$.
The rim that follows most closely the chiral circumferential vector
of a tube has $N_{a}=2m$ armchair and $N_{z}=n-m$ zigzag dangling
bonds. For the $\left(5,5\right)$ armchair tube this rim is shown
in Fig.~\ref{fig:fig1}\emph{a)}. During the growth the total number
of dangling bonds in the rim remains constant $N_{a}+N_{z}=n+m$,
while $N_{a}$ and $N_{z}$ vary. By this condition we include all
reasonable configurations of a growing nanotube and exclude obviously
unreasonable configurations, e.g., one side of the tube being much
longer than the other side.

Following the rims in Figs.~\ref{fig:fig1}\emph{a)}-\emph{d)} from
left to right leads to the notation of the particular configuration
of a rim: \emph{a)} $.aa.aa.aa.aa.aa$, \emph{b)} $.aa.zaaz.aa.aa$,
\emph{c)} $a.aa.aaz.zaa.a$, and \emph{d)} $a.aa.aa.aa.aa.a$ . A
rim thus consists of a combination of $z.z$, $aa.z$, $z.aa$ and
$aa.aa$ growth sites. The $aa.z$ and $z.aa$ growth sites are identical
by symmetry. The $aa$, $za$, $az$, and $zz$ configurations do
not contain a growth site as $\mbox{C}_{2}$ addition does not add
hexagons. Therefore, they do not contribute to the tube elongation
process. In the starting configuration the number of growth sites
in a rim of an $\left(n,m\right)$ nanotube is $N_{aa.aa}=2m-n$, $N_{aa.z}=\min\left(m,n-m\right)$
with {}``$\min$'' the minimum and $N_{z.z}=0$. Note that for tubes
with $2m-n\leq0$ there are only $aa.z$ growth sites and zigzag tubes 
$\left(m=0\right)$ do not contain any growth sites at all. The growth 
of zigzag tubes is supressed in our model and needs an intermediate 
addition of C or $\mbox{C}_{3}$ to induce a growth site for $\mbox{C}_{2}$ 
addition, that we do not consider here. A $\mbox{C}_{2}$
addition to the rim will change the type and the number of growth
sites, see Tab.~\ref{tab:growthsiteoverview}.

\begin{table}
\caption{\label{tab:growthsiteoverview}Overview of the three growth site types
with {}``$\Delta.$'' the change in the growth site number, {}``transition''
the change of the bond structure and examples for the transitions.}

\begin{tabular}{ccccccc}
growth site &  & $\Delta.$ &  & transition &  & example\\
\hline
$aa.aa$ &  & $-1$ &  & $aa.aa\rightarrow zaaz$ &  & Fig.\ref{fig:fig1}\emph{a)$\rightarrow$b)}\\
$aa.z$/$z.aa$ &  & $0$ &  & $aa.z\rightarrow zaa.$/$z.aa\rightarrow.aaz$ &  & Fig.\ref{fig:fig1}\emph{b)$\rightarrow$c)}\\
$z.z$ &  & $+1$ &  & $z.z\rightarrow.aa.$ &  & Fig.\ref{fig:fig1}\emph{c)$\rightarrow$d)}\\
\hline
\end{tabular}

\end{table}

\subsection{C. Growth Factor}
The rim of an $\left(n,m\right)$ nanotube with $n>m>n/2$ can be
divided into a part with a chiral vector $\left(2m-n,2m-n\right)$
that contains $aa.aa$ growth sites and a part with a vector $\left(2n-2m,n-m\right)$
that consists exclusively of $aa.z$ sites. Therefore, all nanotube
rims can be divided in $aa.aa$ containing rim parts and $aa.z$ containing
rim parts. During the growth the number of growth sites contributed
by a rim part containing exclusively $aa.z$ sites remains constant.
The number of growth sites contributed by a rim part containing exclusively
$aa.aa$ sites, however, changes continously during the growth as
is best illustrated by the example of an armchair tube. Growing a
full layer of an armchair rim requires the addition of $2\cdot\left(2m-n\right)=n+m$
carbon dimers. The maximum number of $aa.aa$ growth sites $2m-n$
occurs only at half and full armchair layers. The other $2\cdot\left(2m-n\right)-2$
growth steps have one growth site less%
\footnote{The number of growth sites for other, less likely growth paths will
differ somewhat from Eq.~(\ref{eq:lambdaaa.aa}). However, we verified
Eq.~(\ref{eq:lambdaaa.aa}) to be correct within $\pm5$\% (derived
for the $\left(4,4\right)$ tube in comparison to the next 4 less
likely paths).%
}. Summing up the number of growth sites in each step and dividing
by the number of $\mbox{C}_{2}$ additions yields the average growth
site number

\begin{equation}
\Lambda_{aa.aa}\left(n,m\right)=2m-n-1+\frac{1}{2m-n}.\label{eq:lambdaaa.aa}\end{equation}
Similarily, we find the average growth site number for the rim part
containing $aa.z$ sites $\Lambda_{aa.z}=N_{aa.z}$. Adding the contributions
of $aa.aa$ and $aa.z$ rim parts yields the average number of growth
sites as a function of chiral indexes $n$ and $m$

\begin{equation}
\Lambda\left(n,m\right)=\begin{cases}
\Lambda_{aa.aa}+\Lambda_{aa.z} & \mbox{if }2m-n>0\\
\Lambda_{aa.z} & \mbox{otherwise}.\end{cases}\label{eq:Lambda}\end{equation}

The addition of $\mbox{C}_{2}$ dimers to the $\Lambda\left(n,m\right)$
sites will lead to a lengthening of the tube with $n+m$ $\mbox{C}_{2}$
additions for a single full layer. If we define the abundance of a
certain nanotube chirality to depend on the number of full carbon
layers, we find the growth speed of a tube to be proportional to

\begin{equation}
\Gamma\left(n,m\right)=\frac{\Lambda\left(n,m\right)}{n+m}.\label{eq:Gamma}\end{equation}
The growth factor $\Gamma\left(n,m\right)$ allows us to understand
why chiral selectivity occurs during the nanotube elongation phase.
In the following we will show how we can influence the chiral distribution
during the elongation of a nanotube.

\subsection{D. Influence on Chirality Distribution}
The addition of $\mbox{C}_{2}$ to the different growth sites will
experience varying energy barriers, as zigzag dangling bonds ($E_{z}=2.90\,\mbox{eV}$)
require much more energy than armchair dangling bonds ($E_{a}=2.10\,\mbox{eV}$)
in vacuum~\cite{AndreasThess1996}. The armchair configuration is
energetically favorable because it consists of two dangling bonds
on neighbouring C atoms that form a triple bond. To model experimental
growth conditions we need to consider a metal catalyst in most growth
scenarios. The energetic difference between $a$ and $z$ dangling
bonds is reduced by the presence of a metal~\cite{StephanieReich2006,Yazyev2008,FengDing2009},
as carbon-metal bonds are formed. However, the difference remains
non-zero, as electrons of carbon neighbours influence the total bond
energy of the carbon-metal bonds, rendering $a$ lower in energy than
$z$.

The energy barrier for the $\mbox{C}_{2}$ addition to an $aa.aa$
site depends on the conversion of $aa.aa$ into $zaaz$ dangling bonds
(see Tab.~\ref{tab:growthsiteoverview}). The conversion requires an
energy

\begin{eqnarray}
\Delta_{a} & = & E_{zaaz}-E_{aa.aa}=2E_{z}-2E_{a}=2E_{a}\left(r-1\right),\label{eq:deltaac}\end{eqnarray}
with $E_{a}$ the energy of an armchair and $E_{z}$ the energy of
a zigzag dangling bond. With $r=E_{z}/E_{a}$ we denote the ratio
between the two energies. The total dangling bond energies as well
as their ratio depend on the catalyst. Changing $z.z$ into $.aa.$
we gain $\Delta_{a}$. In contrast growing at an $aa.z$ site will
cost no energy; this growth happens without an energetic barrier.
This energetically different behaviour allows to affect the chirality
distribution of carbon nanotubes through external parameters such
as the metal catalyst and the growth temperature.

\begin{figure}
\includegraphics[scale=0.2]{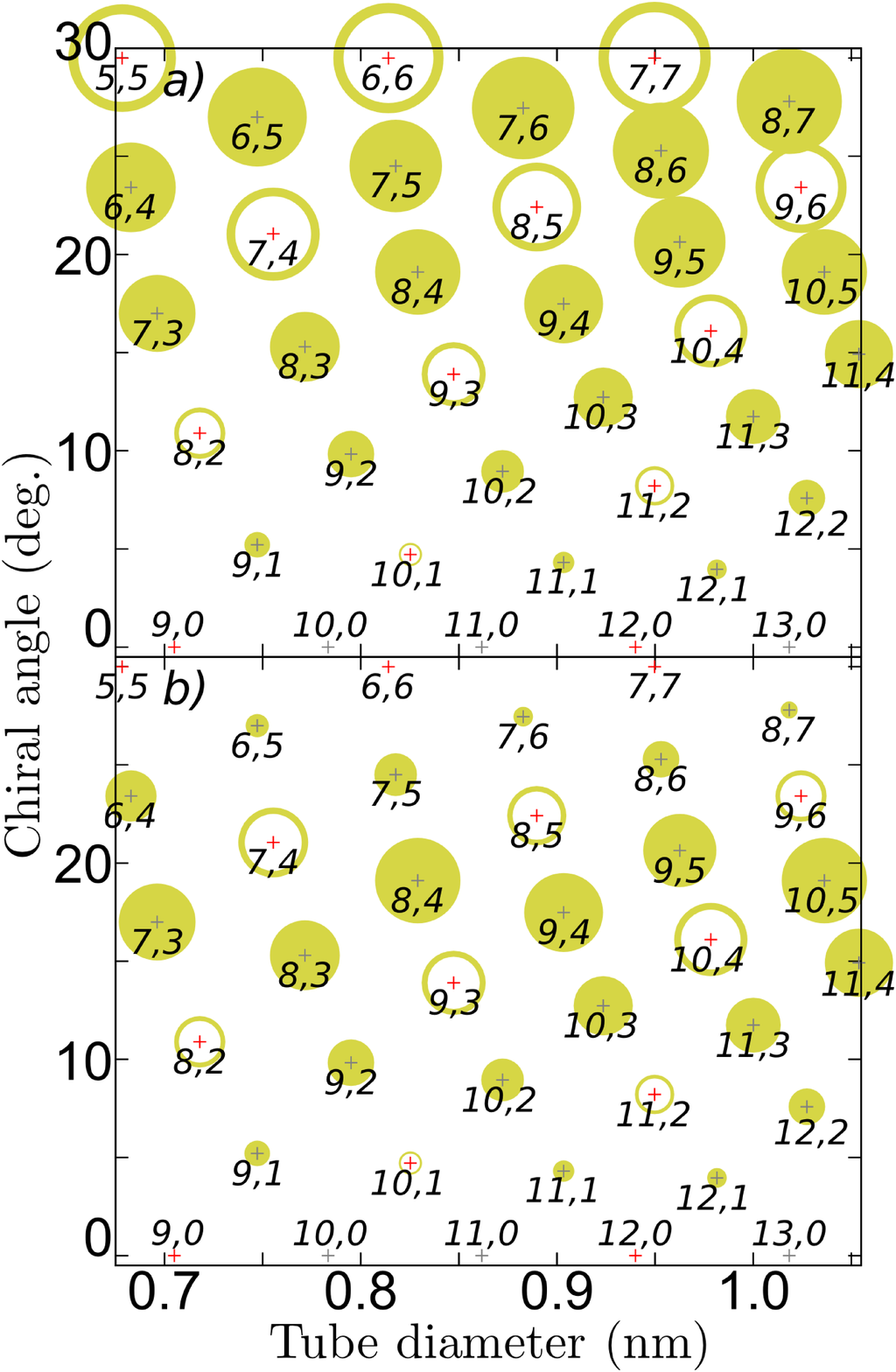}

\caption{\label{fig:gammatheory}Comparison of $\Gamma\left(n,m\right)$ for
tube diameters $d=0.675\mbox{-}1.055\,\text{nm}$ for \emph{a)} $\Delta_{a}\ll k_{B}T$.
\emph{b)} $\Delta_{a}\gg k_{B}T$. The abundance of metallic/semi-metallic
tubes (open circle, red cross) decreases compared to semiconducting
tubes (full circle, gray cross) from \emph{a)} to \emph{b)}.}

\end{figure}

If the addition to $aa.aa$ sites has a negligible barrier $\left(r\approx1\mbox{ or }\Delta_{a}\ll k_{B}T\right)$
all growth sites can contribute to the growth speed. We combine Eqs.~(\ref{eq:Gamma})
and (\ref{eq:Lambda}) to obtain $\Gamma$. Figure \ref{fig:gammatheory}\emph{a)}
shows the growth speed $\Gamma$ as area size in chiral angle and
diameter dependence for diameters $d=0.675\mbox{-}1.055\,\text{nm}$.
The highest $\Gamma$ occur for $\left(n,n\right)$ armchair tubes.
A small trend for increasing $\Gamma$ exists for larger diameter
tubes, resulting from the fractional term of Eq.~(\ref{eq:lambdaaa.aa}),
as the comparison of the armchair tubes shows. Changing the environment
(e.g. another catalyst with another $r$ or adjustment of temperature)
so that $\Delta_{a}\gg k_{B}T$, the $aa.aa$ growth sites will not
contribute anymore; Eq.~(\ref{eq:Gamma}) yields $\Gamma=\Lambda_{aa.z}/\left(n+m\right)$,
which leads to a different growth speed distribution. The highest
$\Gamma$ now occurs for $\left(n,\frac{n}{2}\right)$ chiral tubes,
see Fig.~\ref{fig:gammatheory}\emph{b)}.

For real samples we expect a distribution of growth speed $\Gamma$
to be between the two limiting cases. The thermal energy of nanotube
growth is on the order of $k_{B}T\approx0.05\mbox{-}0.11\mbox{ eV}$~\cite{MircoCantoro2006,LianxiZheng2009}.
$\Delta_{a}$ depends on the catalyst material, its composition and
- less pronounced - on the position of the carbon with respect to
the metal atom. The barriers for metal catalysts are on the order
of $\Delta_{a}\approx0\mbox{-}0.12\mbox{eV}$ for various metals~\cite{StephanieReich2006,FengDing2009}
and thus comparable to the thermal energy. Therefore, the addition
to the $aa.aa$ site is not suppressed. This agrees with the results
of Ding \emph{et al.}~\cite{FengDing2009}, that the barrier for armchair
kink introduction - which corresponds to $\mbox{C}_{2}$ addition
to $aa.aa$ - is negligible. Recently, other materials like Si$\mbox{O}_{2}$
were found to catalyze nanotube growth~\cite{BiluLiu2009}. Further,
bimetallic catalysts contain different barriers and may be extremely
interesting for influencing the chirality distribution~\cite{Wei-HungChiang2009}.

\begin{figure*}
\includegraphics[scale=0.15]{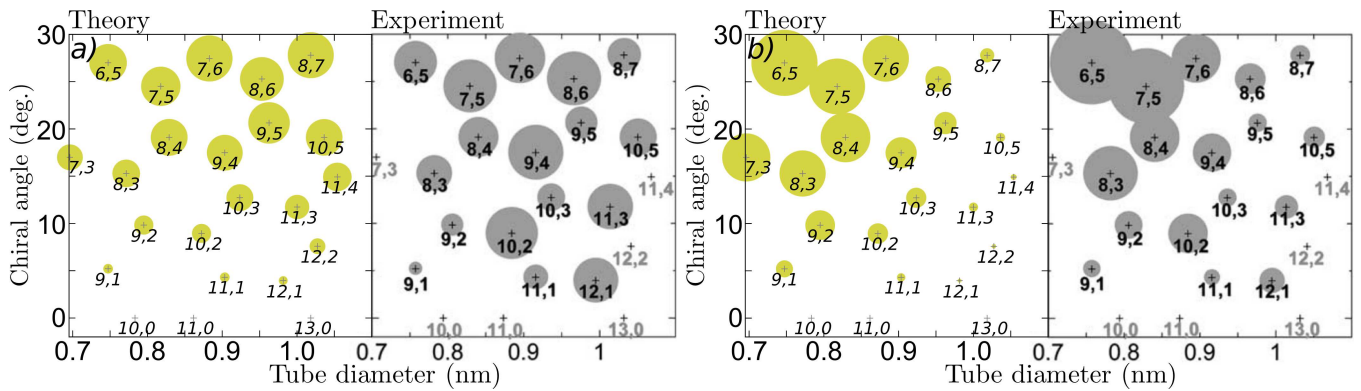}

\caption{\label{fig:fig3}Comparison of chirality distributions. Theoretically
calculated distributions result from a convolution of a gaussian distribution
in diameter dependence and the $\Gamma$ factor for $\Delta_{a}\ll k_{B}T$.
The experimentally determined chirality distributions are adapted
from Miyauchi \emph{et al.}~\cite{YuheiMiyauchi2004}. \emph{a)} Theory:
$d=\left(0.93\pm0.3\right)\,\mbox{nm}$~\cite{SergeiM.Bachilo2002}.
Experiment: HiPco sample. \emph{b)} Theory: $d=\left(0.75\pm0.15\right)\,\mbox{nm}$.
Experiment: ACCVD sample grown at $650\,^{\circ}\mathrm{C}$
with Fe/Co catalyst.}

\end{figure*}

\section{III. Discussion}
Up to now we concentrated on the growth of an existing nanotube nucleus.
The chirality distribution of a sample will also depend on the nucleation
phase, i.e., whether a particular tube cap is nucleated or not~\cite{S.Reich2005}.
The diameter of carbon nanotubes is clearly determined by the nucleation
step~\cite{YimingLi2001}. We now assume a distribution of chiral
indices where (i) the diameter is fixed by nucleation and (ii) the
chiral angle distribution is given by Eq.~(\ref{eq:Gamma}) with $\Delta_{a}\ll k_{B}T$.
Figure \ref{fig:fig3}\emph{a)} compares the chirality distribution
of semiconducting nanotubes with $d=\left(0.93\pm0.3\right)\,\mbox{nm}$
to the experimental distribution in HiPco tubes; Fig.~\ref{fig:fig3}\emph{b)}
is for tubes with $d=\left(0.75\pm0.15\right)\,\mbox{nm}$ and ACCVD
samples~\cite{YuheiMiyauchi2004}. The agreement between theory and
experiment in Fig.~\ref{fig:fig3}\emph{b)} is striking. Our model very 
well predicts the overall decrease of the number of tubes with increasing 
chiral angle. The strong discrepancies for selected chiralities - e.g. 
the strong luminescence of the $\left(10,2\right)$ tube - is most likely 
due to a high quantum yield for some nanotubes~\cite{StephanieReich2005}.
On the other hand, the nucleation phase might also prefer certain
chiralities~\cite{StephanieReich2006}. It would be highly desirable
to establish an unambigious chirality distribution experimentally
to clarify these points. 

Figure~\ref{fig:gammatheory}\emph{a)} verifies our assumption, that a 
mixture of the $\Gamma$ factors derived for $\Delta_{a}\ll k_{B}T$ and 
$\Delta_{a}\gg k_{B}T$ have to be used for real samples as the deviation 
between the theoretical and experimental part of Fig.~\ref{fig:fig3}\emph{a)} show.
Figure~\ref{fig:fig3}\emph{b)} on the other hand perfectly 
reproduces the trend with only considering the contribution of the
$\Gamma$ factor for $\Delta_{a}\ll k_{B}T$, which is expected to 
result from the growth conditions. We conclude that different growth 
conditions have indeed an influence on the chirality distributions which 
results during the elongation of the nanotubes.

\section{IV. Conclusions}
In summary, we suggested how to control the nanotube growth and elongation
process through the structure of the rim. Depending on the tube chirality
the rim contains three different growth sites. Geometric considerations
yield the growth factor $\Gamma$, which in turn determines the chirality
distribution of carbon nanotube samples. We showed that chiral selectivity
can be obtained through a combination of external parameters. Our
results will be important for the understanding and tailoring of the
growth process of single-walled carbon nanotubes.

\section{Acknowledgments}
\begin{acknowledgments}
We acknowledge useful discussions with J. Robertson and S. Heeg.  This work 
was supported by ERC grant no. 210642.
\end{acknowledgments}


\end{document}